\documentstyle[amstex,amssymb,12pt,]{article}

\input{tcilatex}

\begin{document}

\begin{center}
{\LARGE Integral Operators Basic in Random Fields 
Estimation Theory}

\vspace{0.7cm}

{\bf Alexander Kozhevnikov } \noindent \\[0pt]
Department of Mathematics, \\[0pt]
University of Haifa,
Haifa, 31905, ISRAEL \\[0pt]
kogevn@@math.haifa.ac.il
\end{center}

\vspace{0.7cm}

\begin{center}
{\bf Alexander G. Ramm}
Department of Mathematics, \\[0pt]
Kansas State University \\[0pt]
Manhattan, KS 66502, USA \\[0pt]
ramm@@math.ksu.edu
\end{center}

\vspace{0.7cm}

{\bf Abstract}.
\footnote{Math subject classification: 35S15,  35R30,
45B05, 45P05, 62M09, 62M40;\\
key words: integral equations, pseudodifferential operators,
random fields estimation, boundary-value problems, Fredholm property}
The paper deals with the basic integral
equation of random field estimation theory by the
criterion of minimum of variance of the error estimate. This
integral equation is of the first kind. The corresponding integral 
operator over a bounded domain $\Omega $ in ${\Bbb
R}^{n}$ is weakly  singular. This operator is 
an isomorphism between appropriate
Sobolev spaces. This is proved by a reduction of the integral equation to 
an elliptic
boundary value problem in the domain exterior to $\Omega .$
Extra difficulties arise due to the fact that the exterior
boundary value problem should be solved in the Sobolev
spaces of negative order. 

\section{ Introduction}

For convenience of the reader, all standard notations, such
as ${\Bbb N}_{+},\, {\Bbb R},\,{\Bbb C}$, the
definitions of the Fourier transform, of the Sobolev spaces, 
etc, are placed in the Appendix.

Let $P$\ \ be a differential operator in ${\Bbb R}^{n}\;$\
of order $\mu $, 
\ \[ P:=P\left(
x,D\right) :=\sum\limits_{|\alpha |\leq \mu }a_{\alpha
}\left( x\right) D^{\alpha }, \] 
where $a_{\alpha }\left(
x\right) \in C^{\infty }\left( {\Bbb R}^{n}\right) . $

The polynomials 
\[
p\left( x,\xi \right) :=\sum\limits_{|\alpha |\leq \mu }a_{\alpha }\left(
x\right) \xi ^{\alpha }\quad \text{and\quad\ }p_{0}\left( x,\xi \right)
:=\sum\limits_{|\alpha |=\mu }a_{\alpha }\left( x\right) \xi ^{\alpha } 
\]
\ \ are called respectively a symbol and a principal symbol 
of $P.$

Suppose that the symbol $p(x,\xi )$ belongs to the class $SG^{\left( \mu
,0\right) }\left( {\Bbb R}^{n}\right) $ consisting of all $C^{\infty }$
functions $p\left( x,\xi \right) $ on ${\Bbb R}^{n}\times {\Bbb R}^{n}$ that
for any multiindices $\alpha ,\beta $ there exists a constant $C_{\alpha
,\beta }$ such that \ 
\begin{equation}
\left| D_{x}^{\alpha }D_{\xi }^{\beta }p\left( x,\xi \right) \right| \leq
C_{\alpha ,\beta }\left\langle \xi \right\rangle ^{\mu -|\beta
|}\left\langle x\right\rangle ^{-|\alpha |}\quad \left( x,\xi \in {\Bbb R}%
^{n},\quad {\langle \xi \rangle :=}(1+|\xi |^{2})^{1/2}\right)  \label{symb}
\end{equation}

It is known (cf. \cite[Prop. 7.2]{Wloka95}) that the map $P\left( x,D\right)
:{\cal S}\left( {\Bbb R}^{n}\right) \rightarrow {\cal S}\left( {\Bbb R}%
^{n}\right) $ is continuous, where ${\cal S}\left( {\Bbb R}^{n}\right) $ is
the space of \ smooth rapidly decaying functions (see Appendix). Let 
$H^{s}\left( 
{\Bbb R}^{n}\right) $ $\left( s\in {\Bbb R}\right) $ be the usual Sobolev
space (see Appendix). It is known that $P\left( x,D\right) $ naturally acts
on the Sobolev spaces, i.e. the operator $P\left( x,D\right) $ is (cf. \cite[%
Sec. 7.6]{Wloka95}) a bounded operator: $H^{s}\left( {\Bbb R}^{n}\right)
\rightarrow H^{s-\mu }\left( {\Bbb R}^{n}\right) $ for all $s\in {\Bbb R}.$

$P\left( x,D\right) $ is called\ elliptic, if $p_{0}\left( x,\xi \right)
\neq 0$ for any $x\in {\Bbb R}^{n},$ $\xi \in {\Bbb R}^{n}\setminus \{0\}.$

Let $P\left( x,D\right) $ and $Q\left( x,D\right) $ be
both elliptic differential operators of even orders $\mu $ and $\nu $
respectively, $0\leq \mu <\nu ,$ with symbols satisfying (\ref{symb}) (for $%
Q\left( x,D\right) $ we replace $p$ and $\mu $ \ in (\ref{symb})
respectively by $q$ and $\nu $). 
The case $\mu \geq \nu$ is a simpler case which leads to an elliptic
operator perturbed by a compact integral operator  in a bounded domain.

{\it We assume also that $P\left( x,D\right) $
and $Q\left( x,D\right) $ are invertible operators, i.e. there exist the
inverse bounded operators $P^{-1}\left( x,D\right) :$ $H^{s-\mu }\left( 
{\Bbb R}^{n}\right) \rightarrow H^{s}\left( {\Bbb R}^{n}\right) $ and $%
Q^{-1}\left( x,D\right) :$ $H^{s-\nu }\left( {\Bbb R}^{n}\right) \rightarrow
H^{s}\left( {\Bbb R}^{n}\right) $ for all $s\in {\Bbb R}.$}

Let $R:=Q^{-1}\left( x,D\right) P\left( x,D\right) .$ The ellipticity and
invertibility of $P\left( x,D\right) $ and $Q\left( x,D\right) $ imply
that $R$ is an elliptic invertible pseudodifferential operator of negative
order $\mu -\nu $ acting from $H^{s}\left( {\Bbb R}^{n}\right) $ onto $%
H^{s+\nu -\mu }\left( {\Bbb R}^{n}\right) $ \ $\left( s\in {\Bbb R}\right) .$

Since $P$ and $Q$ are elliptic, their orders 
$\mu $ and $\nu $ are even for $%
n>2.$ If $n=2$, we assume that $\mu $ and $\nu $ are even numbers.
Therefore, the number $a:=\left( \nu -\mu \right) /2> 0$ is an integer.

Let $\Omega $ denote a bounded connected open set in ${\Bbb R}^{n}$ with a
smooth boundary $\partial \Omega $ ($C^{\infty }$-class surface) and $%
\overline{\Omega }$ its closure in $L^2(\Omega)$, i.e. $\overline{\Omega 
}=\Omega \cup
\partial \Omega .$ The smoothness restriction on the domain can be 
weakened, but we do not go into detail.

The restriction $R_{\Omega }$ of the operator $R$ to the domain $\Omega
\subset {\Bbb R}^{n}$ is defined as \ 
\begin{equation}
R_{\Omega }:=r_{\Omega }R{}e_{\Omega _{-}},  \label{restr}
\end{equation}
where $e_{\Omega _{-}}$ is the extension by zero to $\Omega _{-}:={\Bbb R}%
^{n}\setminus \overline{\Omega }$ \ and $r_{\Omega }$ is the restriction to $%
\Omega .$

It is known (cf. \cite[Th. 3.11, p. 312]{Gr90}) that the operator $R_{\Omega
}$ defines a continuous mapping 
\[
R_{\Omega }:H^{s}\left( \Omega \right) \rightarrow H^{s+\nu -\mu }\left(
\Omega \right) \;\quad \left( s>-1/2\right), 
\]
where $H^{s}\left( \Omega \right) $ is the spaces of restrictions of
elements of $H^{s}\left( {\Bbb R}^{n}\right) $ to $\Omega $ with the usual
infimum norm (see Appendix).

The pseudodifferential operator $R$ of negative order $\mu -\nu $ as well as
its restriction $R_{\Omega }$ can be represented as integral operators with
kernel $R\left( x,y\right) :$ 
\[
Rh=\int\limits_{{\Bbb R}^{n}}R\left( x,y\right) h\left( y\right) dy,\quad
R_{\Omega }h=\int\limits_{\Omega }R\left( x,y\right) h\left( y\right)
dy\;\left( x\in \Omega \right) , 
\]
where $R\left( x,y\right) \in C^{\infty }\left( {\Bbb R}^{n}\times {\Bbb R}%
^{n}\setminus Diag\right) ,$ $Diag$ is the diagonal in ${\Bbb R}^{n}\times 
{\Bbb R}^{n},$ Moreover, $R\left( x,y\right) $ has a weak singularity: 
\[
\left| R\left( x,y\right) \right| \leq C\left| x-y\right| ^{-\sigma }
\quad n+\mu -\nu <\sigma<n . 
\]
For $n+\mu -\nu <0,$ $R\left( x,y\right) $ is continuous.

Let $\gamma:=n+\mu -\nu$ and $r_{xy}:=|x-y|\to 0$.
Then $R(x,y)=O(r^{-\gamma}_{xy})$ if $n$ is odd or if $n$ is even and 
$\nu<n$, and $R(x,y)=O(r^{-\gamma}_{xy} \log r_{xy})$ if $n$ is even
and $\nu>n$.

In \cite{Ramm1}, the equation 
\begin{equation}
R_{\Omega }h=f\in H^{a}\left( \Omega \right) ,\quad h\in H_{0}^{-a}\left(
\Omega \right)  \label{eq1}
\end{equation}
is derived as a necessary condition for the optimal estimate of random
fields by the criterion of minimum of variance of the error of the 
estimate.
The kernel $R(x,y)$ is a known covariance function, and $h(x,y)$ is the
distributional kernel of the operator of
optimal filter. The kernel $h(x,y)$ should be of minimal order of 
singularity (\cite{Ramm1}). In \cite%
{Ramm1}, the case was considered when $P$ and $Q$ are polynomial functions
of a selfadjoint elliptic operator defined in the whole space. In \cite%
{Ramm2} and \cite{Ramm3}, some generalizations of this theory are obtained.
This paper presents an extension of some results from \cite{Ramm2}.

The purpose of this paper is to prove that, under some natural 
assumptions, the restriction operator $R_{\Omega }$ is an isomorphism of the
space $H_{0}^{-a}\left( \Omega \right) $ onto $H^{a}\left( \Omega \right) $
where $a=\left( \nu -\mu \right) /2>0$ and $H_{0}^{s}\left( \Omega \right) $ 
$\left( s\in {\Bbb R}\right) $ \ denotes the subspace of $H^{s}\left( {\Bbb R%
}^{n}\right) $ that consists of the elements supported in 
$\overline{\Omega }$ (see
Appendix).

To prove the isomorphism property, we reduce the integral equation (\ref{eq1}%
) to an equivalent elliptic exterior boundary-value problem. Since we look
for a solution $u$ belonging to the space $H^{a}\left( \Omega _{-}\right)
=H^{\left( \nu -\mu \right) /2}\left( \Omega _{-}\right) $ and the
differential operator $Q$ is of order $\nu ,$ then $Qu$ should belong to
some Sobolev space of negative order.\ This means that we need results on
the solvability of (\ref{nbvp}) in Sobolev spaces of negative order. Such
spaces as well as solvability in them of elliptic differential boundary
value problems in {\it bounded} domains have been investigated by Ya.
Roitberg \cite{Roit96} and later by V. Kozlov, V. Maz'ya, J. Rossmann \cite%
{KMR97}. The case of pseudodifferential boundary
value problems has been studied in \cite{K01}. A. Erkip and E. Schrohe 
\cite{ErkipSchrohe92}
and E. Schrohe \cite{Schrohe99} have proved the solvability of elliptic
differential and pseudodifferential boundary value problems for unbounded
manifolds and in particular for exterior domains. These solvability results
have been obtained in so-called weighted Sobolev spaces and in particular in
the Sobolev spaces $H^{s}$ of positive order $s$. To obtain the isomorphism
property, the solvability results by A. Erkip and E. Schrohe for exterior
domain should be extended to the weighted Sobolev spaces of negative order.
One can find in Appendix the definition of these spaces 
which is
similar to \cite{Roit96} and \cite{KMR97}.

\section{Reduction of the basic integral equation to a boundary-value problem%
}

{\bf Theorem 1}. {\it The integral equation }(\ref{eq1}) {\it is equivalent
to the following system} (\ref{ubp}), (\ref{eq3}), (\ref{extr}): 
\begin{equation}
\left\{ 
\begin{array}{ll}
Qu=0 & \text{in }\Omega _{-} \\ 
D_{{\bf n}}^{j}u=D_{{\bf n}}^{j}f\; & \text{on }\partial \Omega ,\quad 0\leq
j\leq a-1%
\end{array}
\right.  \label{ubp}
\end{equation}
\begin{equation}
\quad Ph=QF,\quad \quad h\in H_{0}^{-a}\left( \Omega \right) \qquad \qquad
\qquad  \label{eq3}
\end{equation}
{\it where }$u\in H^{a}\left( \Omega _{-}\right) $ {\it is an
extension-function for} $f,$ {\it i.e.} 
\begin{equation}
\qquad \quad \quad F\in H^{a}\left( {\Bbb R}^{n}\right) ,\quad F:=\left\{ 
\begin{array}{ll}
f\in H^{a}\left( \Omega \right) & \text{in }\Omega , \\ 
u\in H^{a}\left( \Omega _{-}\right) & \text{in }\Omega _{-}\text{ }%
\end{array}
\right.  \label{extr}
\end{equation}

{\bf Proof}. Let $h\in H_{0}^{-a}\left( \Omega \right) $ be a solution to (%
\ref{eq1}), i.e. $R_{\Omega }h=f\in H^{a}\left( \Omega \right) $. Let us
define $F:=$ $Q^{-1}Ph.$ Since $h\in H_{0}^{-a}\left( \Omega \right) ,$ then 
$Ph$ $\in H^{-a-\mu }\left( {\Bbb R}^{n}\right) $ and $F=Q^{-1}Ph\in
H^{-a+\nu -\mu }\left( {\Bbb R}^{n}\right) =H^{a}\left( {\Bbb R}^{n}\right)
. $ We have $f=R_{\Omega }h=r_{\Omega }Q^{-1}Ph=r_{\Omega }F,$ i.e. $F$ is
an extension of $f.$ Therefore, $F$ can be represented in the form (\ref%
{extr}). Further, since $F=$ $Q^{-1}Ph,$ then $Ph=QF,$ i.e. $h$ is a
solution to {\bf (}\ref{eq3}{\bf ).} Since $h\in H_{0}^{-a}\left( \Omega
\right) ,$ then $QF=Ph\in H_{0}^{-a-\nu }\left( \Omega \right) .$ It
follows, that $Qu=0$ in $\Omega _{-}.$ Since $F\in H^{a}\left( {\Bbb R}%
^{n}\right) ,\;$we get $D_{{\bf n}}^{j}u=D_{{\bf n}}^{j}f$ on $\partial
\Omega _{-},\quad 0\leq j\leq a-1.$ This means that $\;u\in H^{a}\left(
\Omega _{-}\right) $ is a solution to the boundary value problem (\ref{ubp}%
). Thus, it was proved that any solution to (\ref{eq1}) is also a solution
to the system (\ref{ubp}), {\bf (}\ref{eq3}{\bf ).}

In the opposite direction, let a pair $\left( u,h\right) \in H^{a}\left(
\Omega _{-}\right) \times H_{0}^{-a}\left( \Omega \right) $ be a solution to
the system (\ref{ubp}), {\bf (}\ref{eq3}), (\ref{extr}){\bf . }Since $Ph=QF,$
then $Rh=Q^{-1}Ph=F.$ It follows from (\ref{extr}) that $R_{\Omega }h=\left.
Rh\right| _{\Omega }=\left. F\right| _{\Omega }=f,$ i.e. $h$ is a solution
to (\ref{eq1}). $\blacksquare $\newline

{\bf Remark}. {\it If $\mu >0,$ the boundary value problem (\ref{ubp}) is
underdetermined because $Q$ is an elliptic operator of order $\nu $ and
needs therefore $\nu /2$ boundary conditions, but we have only $a$ $\left(
a<\nu /2\right) $ conditions in (\ref{ubp}). Therefore, the next step is a
transformation of the equation {\bf (}\ref{eq3}{\bf ) }into $\mu /2$ extra
boundary conditions to the boundary value problem (\ref{ubp}). After some
preparations it will be done in Theorem 2.}

{\bf Notation}. Let $\Xi _{+,\lambda }^{t}$\ denote a family $\left( \lambda
\in {\Bbb R},\;t\in {\Bbb Z}\right) $ of order-reducing pseudodifferential
operators $\Xi _{+,\lambda }^{t}:={\cal F}^{-1}\chi _{+}\left( \xi ,\lambda
\right) {\cal F},$ where $\chi _{+}\left( \xi ,\lambda \right) :=\left(
\left( 1+|\xi ^{\prime }|^{2}+\lambda ^{2}\right) ^{1/2}+i\xi _{n}\right)
^{t}$ is the symbol. It is known that the operator $\Xi _{+,\lambda }^{t}$
maps the space ${\cal S}_{0}\left( \overline{{\Bbb R}}_{+}^{n}\right)
:=\left\{ u\in {\cal S}\left( {\Bbb R}^{n}\right) :\text{supp\thinspace }%
u\subset \overline{{\Bbb R}}_{+}^{n}\right\} $ onto itself and has the
following isomorphism properties for $s\in {\Bbb R}$:

\begin{equation}
\Xi _{+,\lambda }^{t}:H^{s}\left( {\Bbb R}^{n}\right) \simeq H^{s-t}\left( 
{\Bbb R}^{n}\right) ,  \label{iso1}
\end{equation}
\begin{equation}
\Xi _{+,\lambda }^{t}:H_{0}^{s}\left( {\Bbb R}_{+}^{n}\right) \simeq
H_{0}^{s-t}\left( {\Bbb R}_{+}^{n}\right) .  \label{iso2}
\end{equation}

Suppose that there exists a real-valued $C^{\infty }$ function \ $\omega $
with nowhere vanishing differential such that $\partial \Omega =\left\{ x\in 
{\Bbb R}^{n}:\omega \left( x\right) =0\right\} $ and $\Omega _{-}=\left\{
x\in {\Bbb R}^{n}:\omega \left( x\right) >0\right\} .$ This means that $%
\Omega _{-}$ is a particular case of a SG-compatible manifold with boundary $%
\partial \Omega $ (cf. \cite[Sect. 1.2]{ErkipSchrohe92}). Let $%
\bigcup\limits_{j=1}^{J}\Omega _{j}$ be a cover of ${\Bbb R}^{n}$ by
finitely many coordinate charts. Let $\left\{ \varphi _{1},...,\varphi
_{J}\right\} $ be a partition of unity and $\left\{ \psi _{1},...,\psi
_{J}\right\} $ be a set of cut-off functions such that (i) supp$\varphi
_{j}, $ supp$\psi _{j}\subseteq \Omega _{j},$ (ii) $\varphi _{j}\psi
_{j}=\varphi _{j},$ (iii) $D^{\alpha }\varphi _{j}\left( x\right) =O\left(
\left\langle x\right\rangle ^{-|\alpha |}\right) ,$ and (iv) $D^{\alpha
}\psi _{j}\left( x\right) =O\left( \left\langle x\right\rangle ^{-|\alpha
|}\right) .$

Let us define a family of pseudodifferential operators depending on a
parameter $\lambda \geq 0$ 
\[
\Lambda _{+}^{t}:=\sum\limits_{j=1}^{J}\psi _{j}\Xi _{+,\lambda }^{t}\varphi
_{j} 
\]
It is known (\cite{Gr96}, \cite{Schrohe99}) that for large enough 
$\lambda ,$
\ the operator $\Lambda _{+}^{t}$ is an isomorphism: 
\[
\Lambda _{+}^{t}:H^{s}\left( {\Bbb R}^{n}\right) \simeq H^{s-t}\left( {\Bbb R%
}^{n}\right) ,\quad s\in {\Bbb R}, 
\]
as well as an isomorphism: 
\begin{equation}
\Lambda _{+}^{t}:H_{0}^{s}\left( \Omega \right) \simeq H_{0}^{s-t}\left(
\Omega \right) ,\quad s\in {\Bbb R}.  \label{iso3}
\end{equation}

{\bf Lemma 1}. {\it Let }$P\left( x,D\right) ${\it \ be an invertible
differential operator of order }$\mu ${\it , i.e. there exists the inverse
operator }$P^{-1}\left( x,D\right) ${\it \ which is bounded: }$H^{s-\mu
}\left( {\Bbb R}^{n}\right) \rightarrow H^{s}\left( {\Bbb R}^{n}\right) $%
{\it \ for all }$s\in R.${\it \ Then a solution }$h${\it \ to the equation } 
\[
P\left( x,D\right) h=g\in H_{0}^{-a-\mu }\left( \Omega \right) 
\]
{\it belongs to the space }$H_{0}^{-a}\left( \Omega \right) ${\it \ if and
only if }$g${\it \ satisfies the following $\mu /2$ boundary conditions:} 
\[
r_{\partial \Omega }D_{{\bf n}}^{j}\Lambda _{+}^{-a-\mu /2}P^{-1}\left(
x,D\right) g=0\quad \left( j=0,...,\mu /2-1\right) . 
\]

{\bf Proof. Necessity}. Let $h=P^{-1}\left( x,D\right) g\in H_{0}^{-a}\left(
\Omega \right) $\ be a solution to $P\left( x,D\right) h=g\in H_{0}^{-a-\mu
}\left( \Omega \right) .{\sf \ }$\ By (\ref{iso3}), we have $\Lambda
_{+}^{-a-\mu /2}h\in H_{0}^{\mu /2}\left( \Omega \right) .$\ Therefore, $%
r_{\partial \Omega }D_{{\bf n}}^{j}\Lambda _{+}^{-a-\mu /2}h=0\quad \left(
j=0,...,\mu /2-1\right) .$\newline

{\bf Sufficiency}. Assume that the equalities $r_{\partial \Omega }D_{{\bf n}%
}^{j}\Lambda _{+}^{-a-\mu /2}h=0\quad \left( j=0,...,\mu /2-1\right) $ hold.
Since $g\in H_{0}^{-a-\mu }\left( \Omega \right) \subset H^{-a-\mu }\left( 
{\Bbb R}^{n}\right) ,$ we have $h=P^{-1}\left( x,D\right) g\in H^{-a}\left( 
{\Bbb R}^{n}\right) .$ Therefore, $\Psi :=\Lambda _{+}^{-a-\mu /2}h\in
H^{\mu /2}\left( {\Bbb R}^{n}\right) .$ Since $r_{\partial \Omega }D_{{\bf n}%
}^{j}\Psi =0\quad \left( j=0,...,\mu /2-1\right) $, we have $\Psi =\Psi
_{+}+\Psi _{-}$, where $\Psi _{+}:=e_{\Omega _{-}}r_{\overline{\Omega }}\Psi
\in H_{0}^{\mu /2}\left( \Omega \right) $ and $\Psi _{-}$ $:=e_{\Omega
}r_{\Omega _{-}}\Psi \in H_{0}^{\mu /2}\left( \Omega _{-}\right) .$ Since $%
\Lambda _{+}^{n/2}:$ $H_{0}^{\mu /2}\left( \Omega \right) \simeq
H_{0}^{-a}\left( \Omega \right) ,\;\Lambda _{+}^{\nu /2}\Psi _{+}\in
H_{0}^{-a}\left( \Omega \right) .$ Moreover, $\Lambda _{+}^{\nu /2}$ is a
differential operator with respect to the variable $x_{n},$ hence supp$\Psi
_{-}\subset \overline{\Omega }_{-}$ implies$\quad $supp$\Lambda _{+}^{\nu
/2}\Psi _{-}\subset \overline{\Omega }_{-}.$ Since $P$ is a differential
operator, 
\[
\text{supp}\left( P\Lambda _{+}^{\nu /2}\right) \Psi _{-}\subset \text{supp}%
\Lambda _{+}^{\nu /2}\Psi _{-}\subset \overline{\Omega }_{-}. 
\]

On the other hand, we have 
\[
\Phi: =\left( P\Lambda _{+}^{\nu /2}\right) \Psi =\left( P\Lambda 
_{+}^{\nu
/2}\right) \left( \Psi _{+}+\Psi _{-}\right) =\left( P\Lambda _{+}^{\nu
/2}\right) \Psi _{+}+\left( P\Lambda _{+}^{\nu /2}\right) \Psi _{-}. 
\]
For any $\varphi \in C_{0}^{\infty }\left( \Omega _{-}\right) $ 
\[
0=\left\langle \Phi ,\varphi \right\rangle =\left\langle \left( P\Lambda
_{+}^{\nu /2}\right) \Psi _{+},\varphi \right\rangle +\left\langle \left(
P\Lambda _{+}^{\nu /2}\right) \Psi _{-},\varphi \right\rangle =\left\langle
\left( P\Lambda _{+}^{\nu /2}\right) \Psi _{-},\varphi \right\rangle 
\]
This means supp$\left( P\Lambda _{+}^{\nu /2}\right) \Psi _{-}\subset 
\overline{\Omega }.$ It follows that supp$\left( P\Lambda _{+}^{\nu
/2}\right) \Psi _{-}\subset \partial \Omega .$ For any $\Psi _{-}\in
C_{0}^{\infty }\left( \Omega _{-}\right) ,$ we have $\left( P\Lambda
_{+}^{\nu /2}\right) \Psi _{-}\in C^{\infty }\left( {\Bbb R}^{n}\right) $
and supp$\left( P\Lambda _{+}^{\nu /2}\right) \Psi _{-}\subset \partial
\Omega .$ $\;$Therefore, $\left( P\Lambda _{+}^{\nu /2}\right) \Psi _{-}=0$
and, moreover, $\Lambda _{+}^{\nu /2}\Psi _{-}=0\quad \left( \forall \Psi
_{-}\in C_{0}^{\infty }\left( \Omega _{-}\right) \right) .$ Since $%
C_{0}^{\infty }\left( \Omega _{-}\right) $ is dense in $H_{0}^{\mu /2}\left(
\Omega _{-}\right) ,$ one gets $\Lambda _{+}^{\nu /2}\Psi _{-}=0\quad \left(
\forall \Psi _{-}\in H_{0}^{\mu /2}\left( \Omega _{-}\right) \right) .$ It
follows, that 
\[
h=\Lambda _{+}^{\nu /2}\Psi =\Lambda _{+}^{\nu /2}\Psi _{+}+\Lambda
_{+}^{\nu /2}\Psi _{-}=\Lambda _{+}^{\nu /2}\Psi _{+}\in H_{0}^{-a}\left(
\Omega \right) . 
\]
$\;\blacksquare $\newline

{\bf Notation.} Let $F\in C^{\infty }\left( \Omega \right) \cap {\cal S}%
\left( \overline{\Omega }_{-}\right) $. Assume that $F$  has finite 
jumps $F_{k}$ of the
normal derivative of order $k$ $\left( k=0,1,...\right) $ on $\partial
\Omega .$ For $x^{\prime }\in \partial \Omega ,$ we will use the following
notation: 
\[
F_{0}\left( x^{\prime }\right) :=\left[ F\right] _{\partial \Omega }\left(
x^{\prime }\right) :=\lim_{\varepsilon \rightarrow +0}\left( F\left(
x^{\prime }+\varepsilon {\bf n}\right) -F\left( x^{\prime }-\varepsilon {\bf %
n}\right) \right) , 
\]
\[
F_{k}\left( x^{\prime }\right) :=\left[ D_{{\bf n}}^{k}F\right] _{\partial
\Omega }\left( x^{\prime }\right) .\qquad \qquad \qquad \qquad \qquad \qquad
\qquad 
\]
Let $f\in C^{\infty }\left( \overline{\Omega }\right) $ and $u\in {\cal S}%
\left( \overline{\Omega }_{-}\right) ,$ then we set $\gamma _{k}f\left(
x^{\prime }\right) :=r_{\partial \Omega }D_{{\bf n}}^{k}f\left( x^{\prime
}\right) ,$ $\gamma _{k}u\left( x^{\prime }\right) :=r_{\partial \Omega }D_{%
{\bf n}}^{k}u\left( x^{\prime }\right) .$

Let $\delta _{\partial \Omega }$ denote the Dirac measure supported on $%
\partial \Omega ,$ i.e. a distribution acting as 
\[
\left( \delta _{\partial \Omega },\varphi \right) :=\int_{\partial \Omega }%
\overline{\varphi }\left( x\right) dS,\quad \varphi \left( x\right) \in
C_{0}^{\infty }\left( {\Bbb R}^{n}\right) . 
\]

It is known that for any differential operator $Q$ of order $\nu $ there
exists a representation $Q=\sum\limits_{j=0}^{\nu }Q_{j}D_{{\bf n}}^{j},\;$%
where $Q_{j}$ is a tangential differential operator of order $\nu -j$ (cf
Appendix). We denote by $\left\{ D^{\alpha }F\left( x\right) \right\} $ the
classical derivative at the points where it exists.

{\bf Lemma 2}. {\it Under the previous notation the following equality holds
for the distribution }$QF${\it :} 
\begin{equation}
QF=\left\{ QF\right\} -i\sum\limits_{j=0}^{\nu
}Q_{j}\sum\limits_{k=0}^{j-1}D_{{\bf n}}^{k}\left( F_{j-1-k}\delta
_{\partial \Omega }\right) .  \label{QF}
\end{equation}

{\bf Proof}. Let $\cos \left( {\bf n}x_{j}\right) $ denote cosine of the
angle between the exterior unit normal vector ${\bf n}$ to the boundary $%
\partial \Omega $ of $\Omega $ and the $x_{j}$-axis.

We use the known formulas 
\[
\int\limits_{\Omega }\frac{\partial u}{\partial x_{j}}dx=\int\limits_{%
\partial \Omega }u\left( x\right) \cos \left( {\bf n}x_{j}\right) d\sigma
\quad u\left( x\right) \in C^{\infty }\left( \overline{\Omega }\right)
,\quad j=1,...,n, 
\]
\[
\int\limits_{\Omega _{-}}\frac{\partial v}{\partial x_{j}}%
dx=-\int\limits_{\partial \Omega }v\left( x\right) \cos \left( {\bf n}%
x_{j}\right) d\sigma \quad v\left( x\right) \in C_{0}^{\infty }\left( 
\overline{\Omega }_{-}\right) \quad j=1,...,n. 
\]
where $d\sigma $ is the surface mesure on $\partial \Omega .$ Applying 
these formulas to the products 
$u\left( x\right) \varphi \left( x\right) $
and $v\left( x\right) \varphi \left( x\right) $ where $\varphi \left(
x\right) \in C_{0}^{\infty }\left( {\Bbb R}^{n}\right) ,$ $u\left( x\right)
\in C^{\infty }\left( \overline{\Omega }\right) ,\;v\left( x\right) \in
C_{0}^{\infty }\left( \overline{\Omega }_{-}\right) ,$ we get 
\begin{equation}
\int\limits_{\Omega }\frac{\partial u}{\partial x_{j}}\varphi \left(
x\right) dx=-\int\limits_{\Omega }u\left( x\right) \frac{\partial \varphi }{%
\partial x_{j}}dx+\int\limits_{\partial \Omega }u\left( x\right) \varphi
\left( x\right) \cos \left( {\bf n}x_{j}\right) d\sigma \quad \quad
j=1,...,n,  \label{Gr1}
\end{equation}
\begin{equation}
\int\limits_{\Omega _{-}}\frac{\partial v}{\partial x_{j}}\varphi \left(
x\right) dx=-\int\limits_{\Omega _{-}}v\left( x\right) \frac{\partial
\varphi }{\partial x_{j}}dx-\int\limits_{\partial \Omega }v\left( x\right)
\varphi \left( x\right) \cos \left( {\bf n}x_{j}\right) d\sigma \quad \quad
j=1,...,n.  \label{Gr2}
\end{equation}
By (\ref{Gr1}), (\ref{Gr2}), we have 
\[
\left( \frac{\partial F}{\partial x_{j}},\varphi \right) =-\left( F,\frac{%
\partial \varphi }{\partial x_{j}}\right) =-\int_{{\Bbb R}^{n}}F\left(
x\right) \frac{\partial \overline{\varphi }\left( x\right) }{\partial x_{j}}%
dx= 
\]
\[
=\int_{{\Bbb R}^{n}}\left\{ \frac{\partial F\left( x\right) }{\partial x_{j}}%
\right\} \overline{\varphi }\left( x\right) dx+\int_{\partial \Omega }\left[
F\right] _{\partial \Omega }\left( x\right) \cos \left( {\bf n}x_{j}\right) 
\overline{\varphi }\left( x\right) dS= 
\]
\[
=\left( \left\{ \frac{\partial F}{\partial x_{j}}\right\} +\left[ F\right]
_{\partial \Omega }\cos \left( {\bf n}x_{j}\right) \delta _{\partial \Omega
},\;\varphi \left( x\right) \right) ,\quad \varphi \left( x\right) \in
C_{0}^{\infty }\left( {\Bbb R}^{n}\right) . 
\]
This means, 
\[
\frac{\partial F}{\partial x_{j}}=\left\{ \frac{\partial F}{\partial x_{j}}%
\right\} +\left[ F\right] _{\partial \Omega }\cos \left( {\bf n}x_{j}\right)
\delta _{\partial \Omega },\quad j=1,...,n. 
\]
It follows, $D_{{\bf n}}F=\left\{ D_{{\bf n}}F\right\} -iF_{0}\delta
_{\partial \Omega }.$ Further, using the last formula we have $D_{{\bf n}%
}^{2}F=D_{{\bf n}}\left\{ D_{{\bf n}}F\right\} -iD_{{\bf n}}\left(
F_{0}\delta _{\partial \Omega }\right) =\left\{ D_{{\bf n}}^{2}F\right\}
-iF_{1}\delta _{\partial \Omega }-iD_{{\bf n}}\left( F_{0}\delta _{\partial
\Omega }\right) $ and so on. By induction one gets: 
\[
D_{{\bf n}}^{j}F=\left\{ D_{{\bf n}}^{j}F\right\}
-i\sum\limits_{k=0}^{j-1}D_{{\bf n}}^{k}\left( F_{j-1-k}\delta _{\partial
\Omega }\right) \quad \left( j=1,2,...\right) . 
\]
Substituting this formula for $D_{{\bf n}}^{j}F$ into the
representation $Q=\sum\limits_{j=0}^{\nu }Q_{j}D_{{\bf n}}^{j},$ we get (\ref%
{QF}). $\blacksquare $

Denoting by $f^{0}$ and $u^{0}$ the extensions by zero to ${\Bbb R}%
^{n}$ of functions $f\left( x\right) \in C^{\infty }\left( \overline{\Omega }%
\right)$ and  $u\left( x\right) \in C_{0}^{\infty }\left( 
\overline{\Omega }%
_{-}\right) ,$ and using Lemma 2, we obtain the following formulas: 
\begin{equation}
\left( Qf\right) ^{0}=Q\left( f^{0}\right) -i\sum\limits_{j=1}^{\nu
}Q_{j}\sum\limits_{k=0}^{j-1}D_{{\bf n}}^{k}\left( \left. \left( D_{{\bf n}%
}^{j-1-k}f\right) \right\vert _{\partial \Omega }\cdot \delta _{\partial
\Omega }\right) \quad \left( f\in C^{\infty }\left( \overline{\Omega }%
\right) \right) ,  \label{Qf0}
\end{equation}%
\begin{equation}
\left( Qu\right) ^{0}=Q\left( u^{0}\right) +i\sum\limits_{j=1}^{\nu
}Q_{j}\sum\limits_{k=0}^{j-1}D_{{\bf n}}^{k}\left( \left. \left( D_{{\bf n}%
}^{j-1-k}u\right) \right\vert _{\partial \Omega }\cdot \delta _{\partial
\Omega }\right) \quad \left( u\in C_{0}^{\infty }\left( \overline{\Omega }%
_{-}\right) \right) .  \label{Qu0}
\end{equation}%
Using these formulas we can define the action of the operator $Q$ 
upon the
elements of the spaces ${\frak H}^{s,\nu }\left( \Omega \right) $ and $%
{\frak H}^{s,\nu }\left( \Omega _{-}\right) $ \ $\left( s\in {\Bbb R}\right) 
$ (defined in Appendix)
 as follows (cf. \cite[Sect. 1.3.2.]{KMR97}, \cite[Sect. 2.4]{Roit96}): 
\begin{equation}
\left( Q\left( f,\underline{\psi }\right) \right) ^{0}:=Q\left( f^{0}\right)
-i\sum\limits_{j=1}^{\nu }Q_{j}\sum\limits_{k=0}^{j-1}D_{{\bf n}}^{k}\left(
\psi _{j-k}\cdot \delta _{\partial \Omega }\right) \quad \left( \left( f,%
\underline{\psi }\right) \in {\frak H}^{s,\nu }\left( \Omega \right) 
\right), 
\label{def1}
\end{equation}%
\begin{equation}
\left( Q\left( u,\underline{\phi }\right) \right) ^{0}:=Q\left( u^{0}\right)
+i\sum\limits_{j=1}^{\nu }Q_{j}\sum\limits_{k=0}^{j-1}D_{{\bf n}}^{k}\left(
\phi _{j-k}\cdot \delta _{\partial \Omega }\right) ,\quad \left( \left( u,%
\underline{\phi }\right) \in {\frak H}^{s,\nu }\left( \Omega _{-}\right)
\right).   \label{def2}
\end{equation}

It is known (\cite{Roit96}) that $Q$ defined respectively in 
(%
\ref{def1}) and (\ref{def2}) is a bounded mapping%
\[
Q:{\frak H}^{s,\nu }\left( \Omega \right) \rightarrow {\cal H}^{s-\nu
}\left( \Omega \right) \text{ and }Q:{\frak H}^{s,\nu }\left( \Omega
_{-}\right) \rightarrow {\cal H}^{s-\nu }\left( \Omega _{-}\right) .
\]%
Moreover, $Q$ is respectively the closure
of the mapping \  $f\rightarrow
Q\left( x,D\right) f$ 
$\quad \left( f\in C^{\infty }\left( \overline{\Omega }%
\right) \right) $ or $u\rightarrow Q\left( x,D\right) u$ $\quad \left( u\in 
{\cal S}\left( \overline{\Omega }_{-}\right) \right) $ between the 
corresponding spaces.

{\bf Notation}. Let $W_{m\ell }$ $\;\left( m=1,...,\mu /2,\quad \ell
=a+1,...,\nu \right) $ be the operator acting on $\partial \Omega $ 
as follows:

\begin{equation}
W_{m\ell }\left( \phi \right) :=i\gamma _{m-1}\Lambda _{+}^{-a-\mu
/2}P^{-1}\sum\limits_{\ell =a+1}^{\nu }\sum\limits_{j=\ell }^{\nu }Q_{j}D_{%
{\bf n}}^{j-\ell }\left( \phi \cdot \delta _{\partial \Omega }\right) ,\quad
\phi \in C^{\infty }\left( \partial \Omega \right).   \label{Wkl}
\end{equation}%
where $\gamma _{k}$ is the restriction to $\partial \Omega $ of the 
$D_{{\bf n}}^{k}$ (cf. Appendix).

The mapping $W_{m\ell }$ is a pseudodifferential operator of order $m-\mu
+\nu /2-1-\ell \;$ acting on $\partial \Omega .$ Therefore, for any real $s,$
it is a bounded operator: 
\[
W_{m\ell }:H^{s}\left( \partial \Omega \right) \rightarrow H^{s-m+\mu -\nu
/2+1+\ell }\left( \partial \Omega \right) .
\]%
For $\left( f,\underline{\psi }\right) \in $ ${\frak H}^{a,\nu }\left(
\Omega \right) ,$ we have $g:=Q\left( f,\underline{\psi }\right) \in
H_{0}^{a-\nu }\left( \Omega \right) ,$ and we set 
\begin{equation}
w_{a+m}:=-\gamma _{m-1}\Lambda _{+}^{-a-\mu /2}P^{-1}g^{0}\quad \left(
m=1,...,\mu /2\right) ,  \label{wak}
\end{equation}%
where the operator $\gamma _{m-1}\Lambda _{+}^{-a-\mu /2}P^{-1}\left(
x,D\right) $ is a trace operator of order $m-1-a-3\mu /2.$ It follows
that 
$ w_{a+m}\in H^{\mu /2-m+1/2}\left( \partial \Omega \right) .$

{\bf Theorem 2. }{\it The integral equation }(\ref{eq1}){\it \ }

\[
R_{\Omega }h=f\in H^{a}\left( \Omega \right) ,\quad h\in H_{0}^{-a}\left(
\Omega \right) 
\]%
{\it is equivalent to the following \ boundary value problem:} 
\begin{equation}
\left\{ 
\begin{array}{ll}
Qu=0 & \text{in }\Omega _{-} \\ 
D_{{\bf n}}^{j}u=D_{{\bf n}}^{j}f\; & \text{on }\partial \Omega ,\quad 0\leq
j\leq a-1 \\ 
\sum\limits_{\ell =a+1}^{\nu }W_{m\ell }\left( \gamma _{\ell -1}u\right)
=w_{a+m} & \text{on }\partial \Omega ,\quad 1\leq m\leq \mu /2%
\end{array}%
\right.   \label{bvp}
\end{equation}%
{\it where the functions }$u,\;f${\it \ and }$h${\it \ are related by the
formulas } 
\[
h=P^{-1}QF,\quad F\in H^{a}\left( {\Bbb R}^{n}\right) ,\quad F:=\left\{ 
\begin{array}{ll}
f\in H^{a}\left( \Omega \right)  & \text{in }\Omega , \\ 
u\in H^{a}\left( \Omega _{-}\right)  & \text{in }\Omega _{-}\text{ }%
\end{array}%
\right. 
\]

{\bf Proof}. Our starting point is Theorem 1. 
Consider
the equation $Ph=QF,\quad \quad h\in H_{0}^{-a}\left( \Omega
\right) .$ Since $F\in H^{a}\left( {\Bbb R}^{n}\right) $ and 
 $Qu=0$ in $\Omega _{-}$ by (\ref{ubp}), then $QF\in H_{0}^{a-\nu }\left( 
{\Bbb R}^{n}\right) =H_{0}^{-a-\mu }\left( {\Bbb R}^{n}\right) .$ 
By Lemma 1, a
solution{\it \ }$h${\it \ }to the equation $Ph=QF\in H_{0}^{-a-\mu }\left(
\Omega \right) $ belongs to the space{\it \ }$H_{0}^{-a}\left( \Omega
\right) ${\it \ }if and only if{\it \ }$QF${\it \ }satisfies the following
$\mu/2$ boundary conditions: 
\begin{equation}
r_{\partial \Omega }D_{{\bf n}}^{m-1}\Lambda _{+}^{-a-\mu /2}P^{-1}QF=0\quad
\left( m=1,...,\mu /2\right) .  \label{extrbc}
\end{equation}%
Since $F=f^{0}+u^{0},$ $QF=Q\left( f^{0}\right) +Q\left( u^{0}\right) .$
Substituting the last expression into (\ref{extrbc}) we have 
\[
\gamma _{m-1}\Lambda _{+}^{-a-\mu /2}P^{-1}Q\left( u^{0}\right) =-\gamma
_{m-1}\Lambda _{+}^{-a-\mu /2}P^{-1}Q\left( f^{0}\right) \quad \left(
m=1,...,\mu /2\right) . 
\]%
In view of (\ref{def1}) and (\ref{def2}), one gets:%
\begin{equation}
\begin{array}{l}
i\gamma _{m-1}\Lambda _{+}^{-a-\mu /2}P^{-1}\sum\limits_{j=1}^{\nu
}Q_{j}\sum\limits_{k=0}^{j-1}D_{{\bf n}}^{k}\left( \phi _{j-k}\cdot \delta
_{\partial \Omega }\right) = \\ 
=\gamma _{m-1}\Lambda _{+}^{-a-\mu /2}P^{-1}\left( Q\left( f,\underline{\psi 
}\right) \right) ^{0}+i\gamma _{m-1}\Lambda _{+}^{-a-\mu
/2}P^{-1}\sum\limits_{j=1}^{\nu }Q_{j}\sum\limits_{k=0}^{j-1}D_{{\bf n}%
}^{k}\left( \psi _{j-k}\cdot \delta _{\partial \Omega }\right)%
\end{array}
\label{matr1}
\end{equation}

Since 
\[
F:=\left\{ 
\begin{array}{ll}
f\in H^{a}\left( \Omega \right) & \text{in }\Omega , \\ 
u\in H^{a}\left( \Omega _{-}\right) & \text{in }\Omega _{-}\text{ }%
\end{array}%
\right. \;\text{ and }F\in H^{a}\left( {\Bbb R}^{n}\right) , 
\]
then $\gamma _{j-1}u=\gamma _{j-1}f\;\;\left( j=1,...,a\right) .$ Therefore, 
$\phi _{j}=\gamma _{j-1}u=\gamma _{j-1}f=\psi _{j}\;\left( j=1,...,a\right)
.\;$

We identify the space $H^{a}\left( \Omega \right) $ with the subspace of $%
{\frak H}^{a,\left( \nu \right) }\left( \Omega \right) $ of all $\left( f,%
\underline{\psi }\right) =\left( f,\psi _{1},...,\psi _{\nu }\right) $ such
that $\psi _{a+1}=...=\psi _{\nu }=0.$ Let $\left( f,\underline{\psi }%
\right) $ belong to this subspace and $\left( u,\underline{\phi }\right)
=\left( u,\phi _{1},...,\phi _{\nu }\right) \in {\frak H}^{a,\left( \nu
\right) }\left( \Omega _{-}\right) .$ Then we can rewrite (\ref{matr1}) as 
\[
\gamma _{m-1}\Lambda _{+}^{-a-\mu /2}P^{-1}\left( Q\left( f,\underline{\psi }%
\right) \right) ^{0}+i\gamma _{m-1}\Lambda _{+}^{-a-\mu
/2}P^{-1}\sum\limits_{j=1}^{\nu }Q_{j}\sum\limits_{\ell =a+1}^{j}D_{{\bf n}%
}^{j-\ell }\left( \phi _{\ell }\cdot \delta _{\partial \Omega }\right) =0.
\]%
Changing the order of summation 
\[
\sum\limits_{j=1}^{\nu }\sum\limits_{\ell =a+1}^{j}=\sum\limits_{\ell
=a+1}^{\nu }\sum\limits_{j=\ell }^{\nu }\,,
\]%
we have%
\begin{equation}
i\gamma _{m-1}\Lambda _{+}^{-a-\mu /2}P^{-1}\sum\limits_{\ell =a+1}^{\nu
}\sum\limits_{j=\ell }^{\nu }Q_{j}D_{{\bf n}}^{j-\ell }\left( \phi _{\ell
}\cdot \delta _{\partial \Omega }\right) =-\gamma _{m-1}\Lambda _{+}^{-a-\mu
/2}P^{-1}\left( Q\left( f,\underline{\psi }\right) \right) ^{0}
\label{equal1}
\end{equation}%
where $m=1,...,\mu /2.$ In view of the notation (\ref{Wkl})\ and (\ref{wak}%
), the equalities (\ref{equal1}) can be rewritten in the form of\ $\mu /2$
equations 
\[
\sum\limits_{\ell =a+1}^{\nu }W_{m\ell }\left( \phi _{\ell }\right)
=w_{a+m}\quad \text{on }\partial \Omega ,\;\;m=1,...,\mu /2.
\]%
Since $\phi _{j}=\gamma _{j-1}u\;$ for $u\in S\left( \overline{\Omega }%
_{-}\right) ,\;$we rewrite these equalities as  
\[
\sum\limits_{\ell =a+1}^{\nu }W_{m\ell }\left( \gamma _{\ell -1}u\right)
=w_{a+m}\quad \text{on }\partial \Omega ,\;\;m=1,...,\mu /2.
\]

These equalities define \ $\mu /2$\ extra boundary 
conditions\ to the
boundary-value problem 
\[
\left\{ 
\begin{array}{ll}
Qu=0\; & \text{in }\Omega _{-}, \\ 
D_{{\bf n}}^{j}u=D_{{\bf n}}^{j}f\; & \text{on }\partial \Omega ,\;\;0\leq
j\leq a-1.%
\end{array}%
\right. 
\]%
$\;\blacksquare $

\section{$\protect\bigskip $Isomorphism property}

We look for a solution $u\in H^{a}\left( \Omega _{-}\right) $\ to the
boundary value problem (\ref{bvp}). Let us consider the following
non-homogeneous boundary value problem associated with (\ref{bvp}): 
\begin{equation}
\left\{ 
\begin{array}{ll}
Qu=w & \text{in }\Omega _{-} \\ 
\gamma _{0}B_{j}u:=\gamma _{0}D_{{\bf n}}^{j-1}u=w_{j}\; & \text{on }%
\partial \Omega ,\quad 1\leq j\leq a \\ 
\gamma _{0}B_{a+m}u:=\sum\limits_{\ell =a+1}^{\nu }W_{m\ell }\left( u_{\ell
-1}\right) =w_{a+m} & \text{on }\partial \Omega ,\quad 1\leq m\leq \mu /2%
\end{array}%
\right.   \label{nbvp}
\end{equation}%
where $w,\;w_{j}\;\left( j=1,...,\nu /2\right) $ are arbitrary elements from
the corresponding Sobolev spaces (see below Theorem 3 and 4)

{\bf Assumption 1}. {\it We assume that the differential operator $Q$ is 
$md$%
-elliptic (cf. \cite[p. 23]{ErkipSchrohe92}), i.e. its symbol $q\left( x,\xi
\right) \in SG^{\left( \nu ,0\right) }\left( {\Bbb R}^{n}\right) $ \ is
invertible for large $|x|+|\xi |$ and 
\[
\left[ q\left( x,\xi \right) \right] ^{-1}=O\left( \left\langle \xi
\right\rangle ^{-\nu }\right) . 
\]}%
Note that the $md$-ellipticity differs from the usual ellipticity which 
says that
the principal symbol $q_{0}\left( x,\xi \right) \neq 0$ for any $\left(
x,\xi \right) \in {\Bbb R}^{n}\times {\Bbb R}^{n}\setminus \{0\}.$

For the formulation of the Shapiro-Lopatinskii condition we need some
notation.

{\bf Notation}. Let $\varepsilon >0$ be a sufficiently small number.
Denote by $U$ ($\varepsilon $-conic neighborhood) the union of all balls 
$B\left(x,\varepsilon \left\langle x\right\rangle \right) ,$ 
with the center $x\in
\partial \Omega $ and radius $\varepsilon \left\langle x\right\rangle .$ Let 
$y=\left( y^{\prime },y_{n}\right) =\left( y_{1},...,y_{n-1},y_{n}\right) $
be normal coordinates in an $\varepsilon $-conic neighborhood $U$ of $%
\partial \Omega ,$ i.e., $\partial \Omega $ may be identified with 
$%
\left\{ y_{n}=0\right\} ,$ $y_{n}$ is the normal coordinate, and the normal
derivative $D_{{\bf n}}$ is $D_{y_{n}}$ near $\partial \Omega .$ Each
differential operator on ${\Bbb R}^{n}$ with $SG$-symbol can be written in $%
U $ as a differential operator with respect to $D_{y^{\prime }}$ and $%
D_{y_{n}} $ within the $SG$-calculus \cite[p. 40]{ErkipSchrohe92}, i.e. 
\[
Q=\sum_{j=0}^{\nu }Q_{j}\left( y,D_{y^{\prime }}\right) D_{y_{n}}^{j} 
\]%
where $Q_{j}\left( y,D_{y^{\prime }}\right) $ are differential operators
with symbols belonging to $SG^{\left( \nu ,0\right) }\left( {\Bbb R}%
^{n}\right) .$ Let 
\[
q\left( y,\xi \right) =q\left( y,\xi ^{\prime },\xi _{n}\right)
=\sum_{j=0}^{\nu }q_{j}\left( y,\xi ^{\prime }\right) \xi _{n}^{j} 
\]%
be the symbol of $Q,$ where $\xi ^{\prime }$ and $\xi _{n}$ are cotangent
variables associated with $y^{\prime }$ and $y_{n}.$

{\bf Assumption 2}. {\it We assume that the operator $Q$ is 
$md$-properly
elliptic (cf. \cite[Assumption 1, p. 40]{ErkipSchrohe92}), i.e. for all
large $|y|+|\xi ^{\prime }|$ the polynomial $q\left( y,\xi ^{\prime
},z\right) $ in the complex variable $z\;$has exactly $\nu /2$ zeroes with
positive imaginary part $\tau _{1}\left( y^{\prime },\xi ^{\prime }\right)
,...,\tau _{\nu /2}\left( y^{\prime },\xi ^{\prime }\right) .$}

We conclude from Assumptions 1 and 2 that the polynomial $q\left( y,\xi
^{\prime },z\right) $ has no real zeroes and it has exactly $\nu /2$ 
zeroes
with negative imaginary part for all large $|y|+|\xi ^{\prime }|.$

In particular, the Laplacian $\Delta $ in the space ${\Bbb R}^{n}$ $\left(
n\geq 2\right) $ is elliptic in the usual sense but not $md$-elliptic, while
the operator $I-\Delta $ is $md$-elliptic as well as $md$-properly elliptic.

Let 
\[
\chi \left( y^{\prime },\xi ^{\prime }\right) :=\left(
1+\sum_{i,j=1}^{n-1}\xi _{i}\left( g\left( y\right) ^{-1}\right) _{ij}\xi
_{j}\right) ^{1/2} 
\]%
where $g=\left( g_{ij}\right) $ is a Riemannian metric on $\partial \Omega .$
We denote%
\[
q^{+}\left( y^{\prime },\xi ^{\prime }\right) :=\prod\limits_{j=1}^{\nu
/2}\left( z-\chi \left( y^{\prime },\xi ^{\prime }\right) ^{-1}\tau
_{j}\left( y^{\prime },\xi ^{\prime }\right) \right) . 
\]

Consider the operators $B_{m}$ $\;\left( m=1,...,\nu /2\right) $ from 
(\ref{nbvp}%
). Each of them is of the form 
\[
B_{m}=\sum_{j=0}^{\nu -1}B_{mj}\left( y^{\prime },D_{y^{\prime }}\right)
D_{y_{n}}^{j} 
\]%
in the normal coordinates $y=\left( y^{\prime },y_{n}\right) =\left(
y_{1},...,y_{n-1},y_{n}\right) $ in an $\varepsilon $-conic neighborhood of $%
\partial \Omega .$ Here $B_{mj}\left( y^{\prime },D_{y^{\prime }}\right) $
is a pseudodifferential operator of order $\rho _{m}-j\;\;\left( \rho
_{m}\in {\Bbb N}\right) $ acting on $\partial \Omega .$ Let $b_{mj}\left(
y^{\prime },\xi ^{\prime }\right) $ denote the principal symbol of $%
B_{mj}\left( y^{\prime },D_{y^{\prime }}\right) .$ Let us note that the
operators $B_{m}$ from the boundary value problem (\ref{nbvp}) are operators
of this type. Scaling down the coefficient operators $B_{mj}$ to order zero,
we set 
\[
b_{m}\left( y^{\prime },\xi ^{\prime },z\right) :=\sum_{j=0}^{\nu
-1}b_{mj}\left( y^{\prime },\xi ^{\prime }\right) \chi \left( y^{\prime
},\xi ^{\prime }\right) ^{-\rho _{m}+j}z^{j} 
\]%
Define polynomials in $z$ 
\[
r_{m}\left( y^{\prime },\xi ^{\prime },z\right) =\sum_{j=0}^{\nu
/2}r_{mj}\left( y^{\prime },\xi ^{\prime }\right) z^{j} 
\]%
as the residues of $b_{m}\left( y^{\prime },\xi ^{\prime },z\right) $ modulo 
$q^{+}\left( y^{\prime },\xi ^{\prime }\right) .$

{\bf Assumption 3}. {\it \ (Shapiro-Lopatinskii condition) The 
determinant 
\newline
$\det \left( \left( r_{mj}\left( y^{\prime },\xi ^{\prime }\right) \right)
_{m,j}\right) $ \ is bounded and bounded away from zero,  i.e. there 
exist
two positive costants }$c$ {\it and} $C${\it \ such that }%
$$
0<c\leq {\it \det \left( \left( r_{mj}\left( y^{\prime },\xi ^{\prime
}\right) \right) _{m,j}\right) \leq C.}
$$
The following Theorem 3 has been proved by K. Erkip and E. Schrohe \cite[Th.
3.1]{ErkipSchrohe92} in essentially more general case of the $SG$-manifold.
The latter includes the exterior of bounded domains which is the easiest
case of the $SG$-manifolds. This case was chosen here just for a simplicity
of the exposition.

{\bf Theorem 3.} (a particular case of \cite[Th. 3.1]{ErkipSchrohe92}, \cite%
{Schrohe99}). {\it If the differential operator }$Q${\it \ \ of even order }$%
\nu ${\it \ satisfies Assumptions 1, 2 and 3, i.e. }$Q${\it \ is }$md${\it %
-elliptic, }$md${\it -properly elliptic and such that the
Shapiro-Lopatinskii condition holds for the operator }$\left( Q,\gamma
_{0}B_{1},...,\gamma _{0}B_{\nu /2}\right) ${\it , then the mapping } 
\[
\left( Q,\gamma _{0}B_{1},...,\gamma _{0}B_{\nu /2}\right) :H^{s}\left(
\Omega _{-}\right) \rightarrow H^{s-\nu }\left( \Omega _{-}\right) \times
\prod\limits_{j=1}^{\nu /2}H^{s-\rho _{j}-1/2}\left( \partial \Omega \right)
\quad \left( s\geq \nu \right) 
\]%
{\it is a Fredholm operator}.

The following Theorem 4 is an extension of Theorem 3 to the case of the
Sobolev spaces of negative order.

{\bf Theorem 4}. {\it Under conditions of Theorem 3, the mapping }$\left(
Q,\gamma _{0}B_{1},...,\gamma _{0}B_{\nu /2}\right) ${\it \ can be extended
to a Fredholm operator } 
\[
\left( Q,\gamma _{0}B_{1},...,\gamma _{0}B_{\nu /2}\right) :{\frak H}^{s,\nu
}\left( \Omega _{-}\right) \rightarrow {\cal H}^{s-\nu }\left( \Omega
_{-}\right) \times \prod\limits_{j=1}^{\nu /2}H^{s-\rho _{j}-1/2}\left(
\partial \Omega \right) \quad \left( s\in {\Bbb R}\right) . 
\]%
{\it \ }

The proof can be obtained just as it was done in \cite{K01} for the
case of a bounded domain. AK hopes to publish the proof
separately\ for a more general case.

{\bf Assumption 4}.{\it \ Suppose that the Fredholm operator $\left( 
Q,\gamma
_{0}B_{1},...,\gamma _{0}B_{\nu /2}\right) $ has the trivial kernel and
cokernel, i.e. it is an invertible operator.}

For example, if the kernel $R(x,y)$ has the property
$(Rh,h)\geq c||h||^2_{H^{-a}_0}$ for all $h\in H^{-a}_0$,
where $c=const>0$ does not depend on $h$, then the operator
in {\it Assumption 4} is invertible (see \cite{Ramm1}).

{\bf Theorem 5}. {\it Under conditions of Theorem 3 and in addition under
Assumption 4, the mapping} $R_{\Omega }$ {\it which is defined in
Introduction} {\it is an isomorphism }$:H_{0}^{-a}\left( \Omega \right)
\rightarrow H^{a}\left( \Omega \right) ${\it .}

{\bf Proof}. Les us consider the operator $\left( Q,\gamma
_{0}B_{1},...,\gamma _{0}B_{\nu /2}\right) $ \ generated by the boundary
value problem (\ref{nbvp}). Taking into account that $\rho _{j}=$order$%
\,B_{j}=j-1\;\left( j=1,...,a\right) $ and $\rho _{j}=$order$\,B_{j}=j-\mu
+\nu /2-2\;\left( j=a+1,...,\nu /2\right) ,$ then the operator is as
follows: 
\[
\left( u,\underline{\phi }\right) \mapsto \left( Q\left( u,\underline{\phi }%
\right) ,\gamma _{0}B_{1}\left( u,\underline{\phi }\right) ,...,\gamma
_{0}B_{\nu /2}\left( u,\underline{\phi }\right) \right) =(w,w_{1},...,w_{\nu
/2})
\]%
It maps the space${\frak H}^{s,\nu }\left( \Omega _{-}\right) $ to the space 
\[
{\cal H}^{s-\nu }\left( \Omega _{-}\right) \times
\prod\limits_{j=1}^{a}H^{s-j+1/2}\left( \partial \Omega \right) \times
\prod\limits_{j=a+1}^{\nu /2}H^{s-j+\mu -\nu /2+3/2}\left( \partial \Omega
\right) \quad \left( s\in {\Bbb R}\right) .
\]

By Theorem 4, the mapping is a Fredholm operator and, in view of Assumption
4, it is an isomorphism. Considering the isomorphism for $s=a$ \  we obtain,
in view of Theorem 2, that the operator $R_{\Omega }$ is an isomorphism of
the space $H_{0}^{-a}\left( \Omega \right) $ onto $H^{a}\left( \Omega
\right) ${\it .} $\blacksquare $

{\bf Example}. Let $P=I$ \ be the identity operator (its order $\mu =0$) and 
$Q=I-\Delta ,$ $\left( \nu =\text{ord\thinspace }Q=2\right) .$ Then, by
Theorem 4, the correspondig operator $R_{\Omega }$ is an isomorphism: $%
H_{0}^{-1}\left( \Omega \right) \rightarrow H^{1}\left( \Omega \right) .$

{\bf Solution of the integral equation} (\ref{eq1}). Under conditions of
Theorem 5 there exists a unique solution to the integral equation (\ref{eq1}%
). The question is how to find this solution. 

Examples of analytic formulas for the solution to the integral equation 
(\ref{eq1}) can be found in \cite{Ramm1}.
The analytical formulas for the solution can be obtained 
only for domains $\Omega$ of special shape, for example, when
$\Omega$ is a ball, and for special operators $Q$ and $P$, for
example, for operators with constant coefficients.

We give such a formula for the solution of equation (\ref{eq1})
assuming $P=I$ and $Q=-\Delta +a^2 I$.
Consider the equation
\begin{equation}
\int_\Omega \frac {\exp (-a|x - y|)}{4\pi |x - y|} h(y)dy 
= f(x), \quad x \in
\overline \Omega \subset {\Bbb R}^3, \quad a > 0, 
\end{equation}
with kernel $R(x,y):=\frac {\exp (-a|x - y|)}{4\pi |x - y|}$, 
$P=I$, and $Q=-\Delta +a^2 I$.  By 
 Theorem 1, one obtains the
unique solution to equation (24) in  $ H_0^{-1}(\Omega)$:
\begin{equation}
h(x) = (-\Delta + a^2)f + \left( \frac {\partial f}
{\partial \bf{n}} 
- \frac{\partial u}{\partial \bf{n}}\right) 
\delta_{\partial\Omega} 
\end{equation}

where  $u$  is the unique solution to the Dirichlet problem in the
exterior domain  $\Omega_-$
\begin{equation}
(-\Delta + a^2)u = 0 \quad\text{in}\quad \Omega_-, \quad 
u|_{\partial\Omega} =
f|_{\partial\Omega},\quad u(\infty)=0,
\end{equation}
$\partial \Omega$  is the boundary of  $\Omega$, 
and
$\delta_{\partial\Omega}$  is the delta function with support  
${\partial\Omega}$.

For any  $\phi \in C_0^\infty ({\Bbb R}^n)$  one has:
$$
\begin{aligned}
\left( (-\Delta + a^2) R,\phi \right) &= 
\left( R, (-\Delta + a^2) \phi
\right) = \int_\Omega f(-\Delta + a^2) \phi dx + 
\int_{\Omega_-} u(-\Delta + a^2)
\phi dx \\
&= \int_\Omega (-\Delta + a^2) f\phi dx + \int_{\Omega_-} 
(-\Delta + a^2)u\phi dx \\
& \qquad - \int_{\partial\Omega} \left( f \frac {\partial \phi}
{\partial \bf{n}} - \frac{\partial f}{\partial \bf{n}} \phi \right) ds + 
\int_{\partial\Omega}\left( u \frac
{\partial \phi}{\partial \bf{n}} - \phi \frac {\partial u}
{\partial \bf{n}}\right)ds \\
&= \int_\Omega (-\Delta + a^2) f \phi dx + 
\int_{\partial\Omega} \left( \frac {\partial
f}{\partial \bf{n}} - \frac {\partial u}{\partial \bf{n}}\right) \phi 
ds, 
\end{aligned}
$$
where the condition  $u = f$  on  $\partial\Omega$  was used.  
Thus, we have checked that formula (25) gives the 
(unique in $H^{-1}_0(\Omega)$ ) solution to equation (24).

\section{Appendix}

{\bf Notation}. We denote by ${\Bbb R}$ the set of real numbers, by ${\Bbb C}
$ the set of complex numbers. Let ${\Bbb Z}:=\{0,\pm 1,\pm 2,...\},$ ${\Bbb N%
}:=\{0,1,...\},$ ${\Bbb N}_{+}:=\{1,2,...\},$ ${\Bbb R}^{n}:=\left\{
x=\left( x_{1},...,x_{n}\right) :x_{i}\in {\Bbb R},\;i=1,...,n\right\} .\;$

Let $\alpha $ be a multi-index, i.e. $\alpha :=({\alpha }_{1},{\ldots },{%
\alpha }_{n}),\ {\alpha }_{j}\in {\Bbb N},\;|\alpha |:={\alpha }_{1}+{\ldots 
}+{\alpha }_{n},$ $i:=\sqrt{-1};$ $D_{j}:=i^{-1}{\partial }/{\partial }{x_{j}%
};$ $D^{\alpha }:={D_{1}^{{\alpha }_{1}}}{D_{2}^{{\alpha }_{2}}}{\ldots }{%
D_{n}^{{\alpha }_{n}}.}$

Let $C^{\infty }\left( \overline{\Omega }\right) $ be the space of
infinitely differentiable up to the boundary functions in $\overline{%
\Omega }.$ Near $\partial \Omega $ there is defined 
A normal vector field ${\bf n}\left( x\right) =\left( n_{1}\left( x\right) ,..
.,n_{n}\left(
x\right) \right) ,$ is defined in a neighborhood of the boundary
$\partial \Omega$ as 
follows: for $x_{0}\in \partial \Omega ,$ ${\bf n}%
\left( x_{0}\right) $ is the unit normal to $\partial \Omega ,$ pointing
into the exterior of $\Omega .$ We set 
\[
{\bf n}\left( x\right) :={\bf n}\left( x_{0}\right) \text{ for }x\text{ of
the form }x=x_{0}+s{\bf n}\left( x_{0}\right) =:\zeta \left( x_{0},s\right) 
\]%
where $x_{0}\in \partial \Omega ,$ $s\in \left( -\delta ,\delta \right) .$
Here $\delta >0$ is taken so small that the representation of $x$ in terms
of $x_{0}\in \partial \Omega $ and $s\in \left( -\delta ,\delta \right) $ is
unique and smooth, i.e., $\zeta $ is bijective and $C^{\infty }$ with $%
C^{\infty }$ inverse, from $\partial \Omega \times \left( -\delta ,\delta
\right) $ to the set $\zeta \left( \partial \Omega \times \left( -\delta
,\delta \right) \right) \subset {\Bbb R}^{n}.$

We call differential operators {\it tangential} when, for $x\in \zeta 
\left(
\partial \Omega \times \left( -\delta ,\delta \right) \right) $, they are 
\
either of the form 
\[
Af=\sum\limits_{j=1}^{n}a_{j}\left( x\right) \frac{{\partial f}}{{\partial }{%
x_{j}}}\left( x\right) +a_{0}\left( x\right) f\text{ \ \ with\ \ }%
\sum\limits_{j=1}^{n}a_{j}\left( x\right) n_{j}\left( x\right) =0,\text{ } 
\]%
or they are products of such operators. The derivative along ${\bf n}$ is 
denoted 
$\partial _{{\bf n}}:$%
\[
\partial _{{\bf n}}f:=\sum\limits_{j=1}^{n}n_{j}\left( x\right) \frac{{%
\partial f}}{{\partial }{x_{j}}}\left( x\right) 
\]%
for $x\in \zeta \left( \partial \Omega \times \left( -\delta ,\delta \right)
\right) .$ Let $D_{{\bf n}}:=i^{-1}\partial _{{\bf n}}.$

Let$\;\Omega _{-}:={\Bbb R}^{n}\setminus \overline{\Omega }$ denote the
exterior of the domain $\Omega ,$\ $\;r_{\partial \Omega },\;r_{\Omega }$ be
respectively the restriction operators to $\partial \Omega ,\;\Omega
:r_{\partial \Omega }f:=\left. f\right| _{\partial \Omega },\;\;r_{\Omega
}f:=\left. f\right| _{\Omega }\;$

Let ${\cal S}\left( {\Bbb R}^{n}\right) $ be the space of \ rapidly
decreasing functions,  that is the space of all $u\in C^{\infty 
}\left( 
{\Bbb R}^{n}\right) $ such that 
\[
\sup_{|\alpha |\leq k}\sup_{x\in {\Bbb R}^{n}}\left\vert \left(
1+|x|^{2}\right) ^{m}D^{\alpha }u\left( x\right) \right\vert <\infty \quad 
\text{for all\ }k,m\in {\Bbb N}. 
\]

Let ${\cal S}\left( \overline{\Omega }_{-}\right) $ be the space of
restrictions of the elements $u\in {\cal S}\left( {\Bbb R}^{n}\right) $ to $%
\overline{\Omega }_{-}$ (this space is equipped with the factor 
topology).

Let $u\in C^{\infty }\left( \overline{\Omega }\right) $ and $v\in {\cal S}%
\left( \overline{\Omega }_{-}\right) ,$ then we set $\gamma
_{k}u:=r_{\partial \Omega }D_{{\bf n}}^{k}u=\left. \left( D_{{\bf n}%
}^{k}u\right) \right| _{\partial \Omega },$ $\gamma _{k}v:=r_{\partial
\Omega }D_{{\bf n}}^{k}v=\left. \left( D_{{\bf n}}^{k}v\right) \right|
_{\partial \Omega }.$

{\bf Definition of the Sobolev spaces}. Let $H^{s}\left( {\Bbb R}^{n}\right) 
$ $\left( s\in {\Bbb R}\right) $ be the usual Sobolev space: 
\[
H^{s}\left( {\Bbb R}^{n}\right) :=\left\{ f\in {\cal S}^{\prime }\;|\;{\cal F%
}^{-1}\left( 1+|\xi |^{2}\right) ^{s/2}{\cal F}f\in L_{2}\left( {\Bbb R}%
^{n}\right) \right\} , 
\]%
\[
\left\Vert f\right\Vert _{H^{s}\left( {\Bbb R}^{n}\right) }:=\left\Vert 
{\cal F}^{-1}\left( 1+|\xi |^{2}\right) ^{s/2}{\cal F}f\right\Vert
_{L_{2}\left( {\Bbb R}^{n}\right) }, 
\]%
where ${\cal F}$ denotes the Fourier transform $f\mapsto {\cal F}_{x{%
\rightarrow }\xi }f(x)=\int_{{\Bbb R}^{n}}e^{-ix\xi }f(x)dx,$ ${\cal F}^{-1}$
its inverse and ${\cal S}^{\prime }={\cal S}^{\prime }\left( {\Bbb R}%
^{n}\right) $ denote the space of tempered distributions which is dual to
the space ${\cal S}\left( {\Bbb R}^{n}\right) .$

Let $H^{s}\left( \Omega \right) $ and $H^{s}\left( \Omega _{-}\right) $ $%
\left( 0\leq s\in {\Bbb R}\right) $ be respectively the spaces of
restrictions of elements of $H^{s}\left( {\Bbb R}^{n}\right) $ to $\Omega $
and $\Omega _{-}.$ The norms in the spaces $H^{s}\left( \Omega \right) $ and 
$H^{s}\left( \Omega _{-}\right) $ are defined by the relations 
\[
\left\Vert f\right\Vert _{H^{s}\left( \Omega \right) }:=\inf \left\Vert
g\right\Vert _{H^{s}\left( {\Bbb R}^{n}\right) }\quad \left( s\geq 0\right)
, 
\]%
\[
\left\Vert f\right\Vert _{H^{s}\left( \Omega _{-}\right) }:=\inf \left\Vert
g\right\Vert _{H^{s}\left( {\Bbb R}^{n}\right) }\quad \left( s\geq 0\right)
, 
\]%
where infimum is taken over all elements $g\in H^{s}\left( {\Bbb R}%
^{n}\right) $ which are equal to $f$ in $\Omega $ respectively in $\Omega
_{-}.$

By $H_{0}^{s}\left( \Omega \right) $ $\left( s\in {\Bbb R}\right) $ \ and $%
H_{0}^{s}\left( \Omega _{-}\right) ,$ we denote the closed subspaces of the
space $H^{s}\left( {\Bbb R}^{n}\right) $ which consist of the elements with
supports respectively in $\overline{\Omega }$ or in $\overline{\Omega }_{-}$
i.e. 
\[
H_{0}^{s}\left( \Omega \right) :=\left\{ f\in H^{s}\left( {\Bbb R}%
^{n}\right) :\text{supp\thinspace }f\subseteq \overline{\Omega }\right\}
\subset H^{s}\left( {\Bbb R}^{n}\right) \quad \left( s\in {\Bbb R}\right) , 
\]%
\[
H_{0}^{s}\left( \Omega _{-}\right) :=\left\{ f\in H^{s}\left( {\Bbb R}%
^{n}\right) :\text{supp\thinspace }f\subseteq \overline{\Omega }_{-}\right\}
\subset H^{s}\left( {\Bbb R}^{n}\right) \quad \left( s\in {\Bbb R}\right) . 
\]%
We define the spaces 
\[
{\cal H}^{s}\left( \Omega \right) :=\left\{ 
\begin{array}{ll}
H^{s}\left( \Omega \right) & s>0 \\ 
H_{0}^{s}\left( \Omega \right) & s\leq 0,%
\end{array}%
\right. 
\]%
\[
{\cal H}^{s}\left( \Omega _{-}\right) :=\left\{ 
\begin{array}{ll}
H^{s}\left( \Omega _{-}\right) & s>0, \\ 
H_{0}^{s}\left( \Omega _{-}\right) & s\leq 0.%
\end{array}%
\right. 
\]

For $s\neq k+1/2\;\left( k=0,1,...,\ell -1\right) ,$ we define the spaces $%
{\frak H}^{s,\ell }\left( \Omega \right) $ and ${\frak H}^{s,\ell }\left(
\Omega _{-}\right) $ respectively as the sets of all 
\[
\left( u,\underline{\phi }\right) =\left( u,\phi _{1},...,\phi _{\ell
}\right) \;\text{and }\left( v,\underline{\psi }\right) =\left( v,\psi
_{1},...,\psi _{\ell }\right) \text{ } 
\]
where $u\in {\cal H}^{s}\left( \Omega \right) ,\;v\in {\cal H}^{s}\left(
\Omega _{-}\right) ,\;\;\underline{\phi }=\left( \phi _{1},...,\phi _{\ell
}\right) $ and $\underline{\psi }=\left( \psi _{1},...,\psi _{\ell }\right) $
are vectors in $\prod\limits_{j=1}^{\ell }H^{s-j+1/2}\left( \partial \Omega
\right) $ satisfying the condition 
\[
\phi _{j}=\left. D_{{\bf n}}^{j-1}u\right| _{\partial \Omega },\quad \psi
_{j}=\left. D_{{\bf n}}^{j-1}v\right| _{\partial \Omega }\quad \quad \text{%
for }j<\min \left( s,\ell \right) . 
\]
The norms in ${\frak H}^{s,\ell }\left( \Omega \right) $ and ${\frak H}%
^{s,\ell }\left( \Omega _{-}\right) $ can be defined as 
\[
\left\| \left( u,\underline{\phi }\right) \right\| _{{\frak H}^{s,\ell
}\left( \Omega \right) }^{2}=\left\| u\right\| _{{\cal H}^{s}\left( \Omega
\right) }^{2}+\sum\limits_{j=1}^{\ell }\left\| \phi _{j}\right\|
_{H^{s-j+1/2}\left( \partial \Omega \right) }^{2}, 
\]
\[
\left\| \left( v,\underline{\psi }\right) \right\| _{{\frak H}^{s,\ell
}\left( \Omega \right) }^{2}=\left\| v\right\| _{{\cal H}^{s}\left( \Omega
\right) }^{2}+\sum\limits_{j=0}^{\ell}\left\| \psi _{j}\right\|
_{H^{s-j+1/2}\left( \partial \Omega \right) }^{2}. 
\]
Since only the components $\phi _{j}$ and $\psi _{j}$ with index $j<s$ can
be chosen independently of $u,$ we can identify ${\frak H}^{s,\ell }\left(
\Omega \right) $ and ${\frak H}^{s,\ell }\left( \Omega _{-}\right) $ with
the following spaces.

For $s,s_{1}\neq k+1/2\;\left( k=0,1,...,\ell
-1\right) ,$
\[
{\frak H}^{s,\ell }\left( \Omega \right) =\left\{
\begin{array}{l}
{\cal H}^{s}\left( \Omega \right) ,\;\ell =0, \\
{\cal H}^{s}\left( \Omega \right) ,\;1\leq \ell <s+1/2, \\
{\cal H}^{s}\left( \Omega \right) \times \prod\limits_{j=[s+1/2]+1}^{\ell
}H^{s-j+1/2}\left( \partial \Omega \right) ,\;0<\left[ 
s+\frac{1}{2}\right]
<\ell , \\
{\cal H}^{s}\left( \Omega \right) \times \prod\limits_{j=1}^{\ell
}H^{s-j+1/2}\left( \partial \Omega \right) ,\;s<\frac{1}{2}%
\end{array}%
\right.
\]%
\[
{\frak H}^{s,\ell }\left( \Omega _{-}\right) =\left\{
\begin{array}{l}
{\cal H}^{s}\left( \Omega _{-}\right) ,\;\ell =0, \\
{\cal H}^{s}\left( \Omega _{-}\right) ,\;1\leq \ell <s+1/2, \\
{\cal H}^{s}\left( \Omega _{-}\right) \times
\prod\limits_{j=[s+1/2]+1}^{\ell }H^{s-j+1/2}\left( \partial \Omega 
\right)
,\;0<\left[ s+\frac{1}{2}\right] <\ell , \\
{\cal H}^{s}\left( \Omega _{-}\right) \times \prod\limits_{j=1}^{\ell
}H^{s-j+1/2}\left( \partial \Omega \right) ,\;s<\frac{1}{2}%
\end{array}%
\right.
\]%
Finally, \ for $s=k+1/2\;\left( k=0,1,...,\ell -1\right) ,$ we define the
spaces ${\frak H}^{s,\ell }\left( \Omega \right) ,$ ${\frak H}^{s,\ell
}\left( \Omega _{-}\right) $ by the method of complex interpolation.

Let us note that for $s\neq k+1/2\;\left( k=0,1,...,\ell -1\right) ,$ the
spaces ${\frak H}^{s,\ell }\left( \Omega \right) ,$ ${\frak H}^{s,\ell
}\left( \Omega _{-}\right) $ are completion of $C^{\infty }\left( 
\overline{%
\Omega }\right) ,$ ${\cal S}\left( \overline{\Omega }_{-}\right) $
respectively in the norms
\[
\left\Vert \left( u,\gamma _{0}u,...,\gamma _{\ell -1}u\right) \right\Vert 
_{%
{\frak H}^{s,\ell }\left( \Omega \right) }^{2}=\left\Vert u\right\Vert _{%
{\cal H}^{s}\left( \Omega \right) }^{2}+\sum\limits_{j=0}^{\ell
-1}\left\Vert \gamma _{j}u\right\Vert _{H^{s-j-1/2}\left( \partial \Omega
\right) }^{2},
\]%
\[
\left\Vert v,\gamma _{0}v,...,\gamma _{\ell -1}v\right\Vert _{{\frak H}%
^{s,\ell }\left( \Omega _{-}\right) }^{2}=\left\Vert v\right\Vert _{{\cal 
H}%
^{s}\left( \Omega _{-}\right) }^{2}+\sum\limits_{j=0}^{\ell -1}\left\Vert
\gamma _{j}v\right\Vert _{H^{s-j-1/2}\left( \partial \Omega \right) }^{2}.
\]


\begin{thebibliography}{99}
\bibitem{ErkipSchrohe92} A. Erkip, E. Schrohe, Normal Solvability of
Elliptic Boundary Value Problems on Asymptotically Flat Manifolds, J. of
Functional Analysis 109 (1992), 22--51.

\bibitem{Gr90} G. Grubb, Pseudo-differential problems in $L_{p}$ spaces,
Commun. in Partial Differ. Eq.,15(3) (1990), 289--340. \ 

\bibitem{Gr96} G. Grubb, Functional Calculus of Pseudodifferential Boundary
Problems, Birkh\"{a}user, Boston-Basel-Berlin, 1996.

\bibitem{K01} {\small \ }A. Kozhevnikov, Complete Scale of Isomorphisms for
Elliptic Pseudodifferential Boundary-Value Problems, J. London Math. Soc.
(2) 64 (2001), 409--422.

\bibitem{KMR97} V. Kozlov, V. Maz'ya, J. Rossmann, Elliptic Boundary Value
Problems in Domains with Point Singularities, AMS, Providence 1997.

\bibitem{Ramm1} A. G. Ramm, Random Fields Estimation Theory, Longman/Wiley,
New York, 1990. {\small \ }

\bibitem{Ramm2} A. G. Ramm, Analytical solution of a new class of integral
equations, Diff. Integral Eqs, 16, N2, (2003), 231-240.

\bibitem{Ramm3} A. G. Ramm, Estimation of random fields, Theory of
Probability and Math. Statistics, 66, (2002), 95-108.

\bibitem{Roit96} Ya. A. Roitberg, Elliptic boundary-value problems in the
spaces of distributions, Kluwer, Dordrecht, 1996.

\bibitem{Schrohe87} E. Schrohe, Spaces of weighted symbols and weighted
Sobolev spaces on manifolds, In: Pseudo-Differential operators, Cordes,
H.O., Gramsch, B., and Widom, H. (eds.) Springer LN Math. 1256, pp. 360-377,
Springer-Verlag, Berlin, 1987. {\small \ }

\bibitem{Schrohe99} E. Schrohe, Frechet Algebra Techniques for Boundary
Value Problems on Noncompact Manifolds, Math. Nachr. 199 (1999), 145--185.

\bibitem{Wloka87} {\small \ }J.T. Wloka, Partial differential equations,
Cambridge University Press, 1987.

\bibitem{Wloka95} {\small \ }J.T. Wloka, B. Rowley, B. Lawruk, Boundary
value problems for elliptic systems, Cambridge University Press, 1995.%
{\small \ }
\end{thebibliography}
\end{document}